\RequirePackage{fix-cm}
\documentclass[smallextended]{svjour3}       
\smartqed  
\usepackage{graphicx}
\usepackage{hyperref}
\usepackage{amsmath,amssymb,cite}
\usepackage{color}

\newcommand{\mean}[1]{\left<#1\right>}
\newcommand{\means}[1]{\langle#1\rangle}

\begin{document}

\title{Quasiperiodicity and valence fluctuation in the spin-1/2 Falicov-Kimball model}
\author{Joji Nasu \and Ryu Shinzaki \and Akihisa Koga}
\institute{J. Nasu \and R. Shinzaki \and A. Koga \at
  Department of Physics, Tokyo Institute of Technology, Meguro, Tokyo 152-8551, Japan
}
\date{Received: date / Accepted: date}

\maketitle

\begin{abstract}
We study the spin-1/2 Falicov-Kimball model with conduction and localized $f$ electrons on the Penrose lattice using the real-space dynamical mean-field theory.
By changing the $f$ electron level, the $f$ electron density at each site
changes continuously, in contrast to periodic systems with first-order valence transitions.
In the intermediate valence regime,
the local $f$ electron number strongly depends on a wider range of the Penrose structure
surrounding its lattice site,
in spite of the local interaction between the conduction and $f$ electrons.
The temperature dependence of the magnetic response is also discussed.
\end{abstract}

\section{Introduction}
\label{intro}

Quasiperiodic systems have attracted considerable attention
since the discovery of the quasicrystals~\cite{Shechtman}.
In the metallic alloys, the quasiperiodic structure disturbs
the formation of the Bloch wave,
which should lead to remarkable peculiarities of electronic and thermal conductivities.
Recently, interesting low temperature properties have been observed in the ytterbium alloys.
In the quasicrystal Au$_{51}$Al$_{34}$Yb$_{15}$, the quantum critical behavior was observed
in the susceptibility and specific heat, while the heavy fermion behavior was observed
in its approximants~\cite{deguchi2012,Matsukawa2014}.
Furthermore, superconductivity was reported in approximants of the related materials~\cite{deguchi2015},
which stimulates further investigations on electron correlations on the quasicrystals.
In these materials, $f$ electrons of Yb ions are considered to be in intermediate valence regime, and hence valence fluctuations should play a role for these exotic phenomena, in addition to the quasiperiodicity.
To clarify the nature of the critical phenomena and superconductivity, several theoretical studies have been done in the viewpoints of correlation effects within a Tsai-Type cluster~\cite{Watanabe2013,Watanabe2015}, disorder effect~\cite{Andrade2015,Otsuki2016}, and effect of quasiperiodic structure ~\cite{Takemori2015,Takemura2015,Shinzaki2016,Sakai2017}.
However, low-temperature properties in quasiperiodic systems with strong electron correlations and valence fluctuations remain unclear.

In this study, to address the many-body effect and valence fluctuation in quasiperiodic systems, we introduce a spin-1/2 Falicov-Kimball model (FKM)~\cite{Falicov1969,PhysRevB.57.11955,Freericks1998,Freericks_rev2003}
on a two dimensional Penrose lattice as a minimal model.
We analyze the finite temperature properties of this model
using the real-space dynamical mean-field theory (RDMFT)~\cite{Georges1996,pruschke1995,muller1989}.
We have confirmed the suppression of valence transitions by changing the $f$ electron level in contrast to the FKM on the Bethe lattice.
At high temperatures, the occupancy of localized $f$ electrons hardly depend on sites due to strong thermal fluctuations.
While decreasing temperature, one can clearly see the site dependence in the $f$ electron number, which is classified by its coordination number.
Further decrease of temperature gives rise to the $f$ electron distribution reflecting the wider range of the crystal structure beyond the neighboring one in the intermediate valence regime.
This structure disappears at the lowest temperature.
We also find that the magnetic susceptibility shows a peculiar temperature dependence in the competing region between the intermediate and commensurate valence regimes.

This paper is constructed as follows. In Sec.~\ref{sec:model-method}, we introduce the spin-1/2 FKM and RDMFT as the calculation method.
In Sec.~\ref{sec:distr-f-electr}, we present the numerical results for the localized $f$ electron density.
The temperature dependence of the magnetic susceptibility is shown in Sec.~\ref{sec:temp-depend-susc}.
Section~\ref{sec:summary} is devoted to the summary.

\section{Model and method}\label{sec:model-method}

\begin{figure}[t]
 \begin{center}
  \includegraphics[width=0.6\columnwidth,clip]{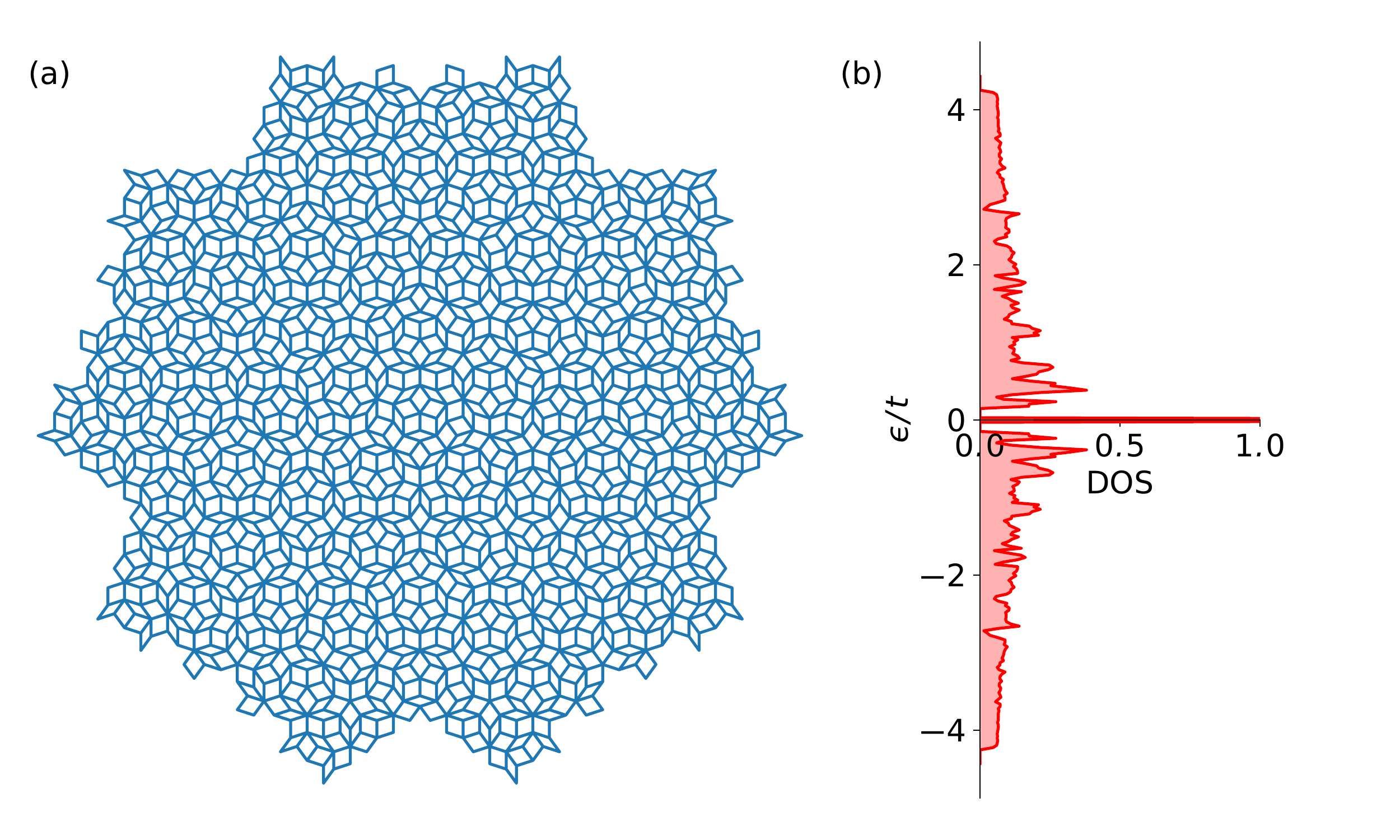}
  \caption{
(a) Penrose lattice with $N_s=1591$ sites used in the present study.
(b) DOS of the tight-binding Hamiltonian ${\cal H}_t$ on the penrose lattice.
}
  \label{penrose}
 \end{center}
\end{figure}

We study the $S=1/2$ FKM on the Penrose lattice including $N_s=1591$ sites shown in Fig.~\ref{penrose}.
The Hamiltonian of this model~\cite{Falicov1969} is given by
\begin{align}
 {\cal H}=-t\sum_{\means{ij}\sigma}
\left(c_{i\sigma}^\dagger c_{j\sigma}
+{\rm H.c.}\right)
+\varepsilon_f\sum_i n_i^f
+U'\sum_{i}n_i^c n_i^f
+U\sum_{i}n_{i\uparrow}^f n_{i\downarrow}^f,
\label{eq:1}
\end{align}
where $c_{i\sigma}$ is the annihilation operator of the conduction electron at site $i$ with spin $\sigma=\uparrow, \downarrow$ and $n_{i\sigma}^c=c_{i\sigma}^\dagger c_{i\sigma}$.
$n_{i\sigma}^f$ is the number operator of the $f$ electron at site $i$ with spin $\sigma$ and $n_i^{c(f)}=n_{i\uparrow}^{c(f)}+n_{i\downarrow}^{c(f)}$.
In Eq.~(\ref{eq:1}), the first term represents the hoppings of the conduction electrons with the transfer integral $t$ between nearest neighbor sites $\means{ij}$ on the Penrose lattice.
The second, third, and fourth terms represent the $f$ electron level $\varepsilon_f$, local Coulomb interaction $U'$ between the conduction and $f$ electrons, and on-site Coulomb interaction $U$ in the $f$ electron orbital.
In the present study, we impose $U\to\infty$,
indicating that the doubly occupied state for $f$ electrons is excluded.
Total electron number $N=\sum_i(n_i^c+n_i^f)$ is fixed to $n=\means{N}/N_s=1.9$ to avoid the appearance of the charge ordered phase.

\begin{figure}[t]
 \begin{center}
  \includegraphics[width=0.8\columnwidth,clip]{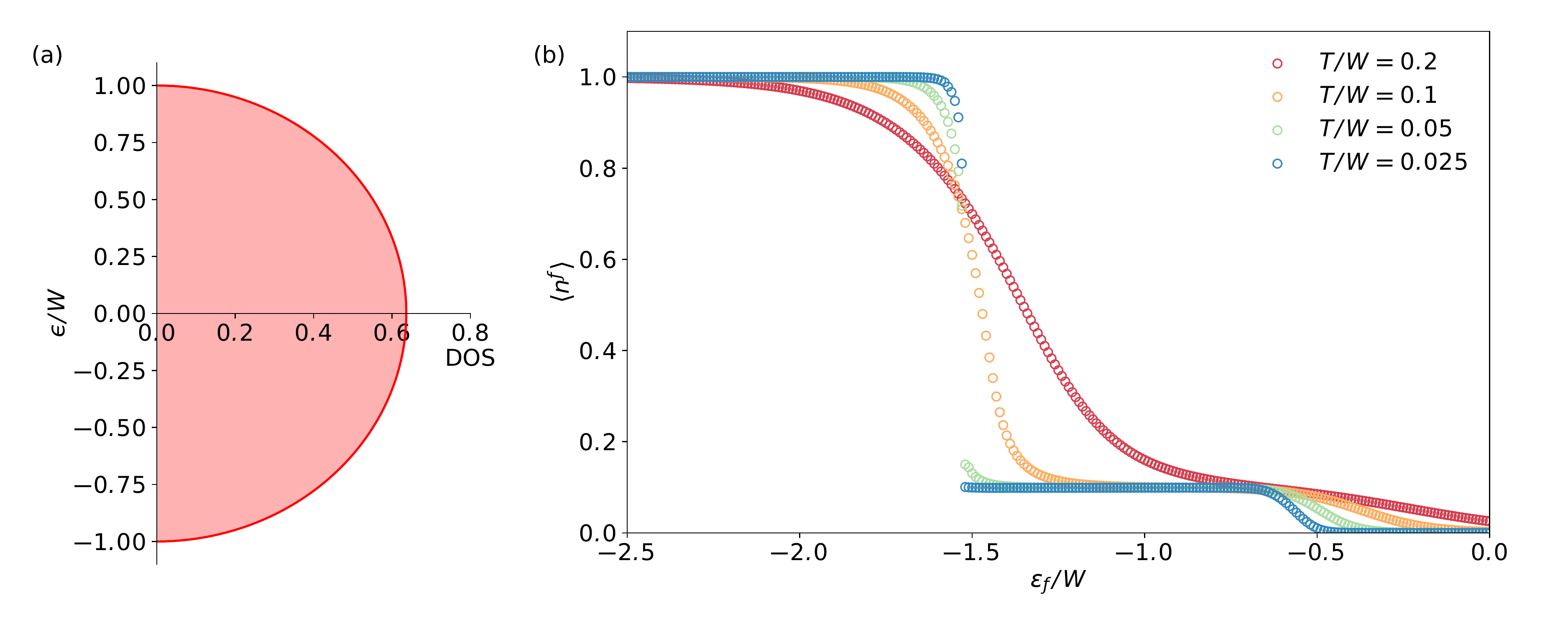}
  \caption{
(a) DOS of the tight binding Hamiltonian on the Bethe lattice.
(b) $f$ electron number as a function of $\varepsilon_f$ in the $S=1/2$ FKM on the Bethe lattice at $U'/W=2$.
}
  \label{bethe}
 \end{center}
\end{figure}


To analyze the model, we adopt the RDMFT~\cite{Georges1996,pruschke1995,muller1989,Metzner1989}.
In this method, the site-dependence of the self-energy $\Sigma$ is taken into account but its inter-site components are neglected.
The noninteracting Green function is given by ${\cal G}_i(i\omega_n)=[G_{ii}^{\rm latt}(i\omega_n)^{-1}+\Sigma_i(i\omega_n)]^{-1}$ in the DMFT scheme, which maps the lattice model onto impurity problems.
Here, $G_{ii}^{\rm latt}$ is the lattice Green function and $\omega_n=(2n+1)\pi T$ is the Matsubara frequency with integer $n$.
An impurity solver updates $\Sigma_i$ using ${\cal G}_i$ for each site.
Finally, the lattice Green function is updated so that $G^{\rm latt}(i\omega_n)^{-1}=(i\omega_n+\mu)I-\Sigma-{\cal H}_t$, where $I$ is the identity matrix and ${\cal H}_t$ is the $N_s\times N_s$ matrix with respect to site representing the first term of Eq.~(\ref{eq:1}).

In general, numerical techniques such as
the continuous-time quantum Monte Carlo method~\cite{CTQMCREV}
are needed to solve impurity problems in DMFT.
Therefore, it is hard to discuss very low temperature properties.
In particular, in the $f$-electron systems on the quasiperiodic structure,
the complex temperature dependence should be expected~\cite{Takemura2015,Shinzaki2016}.
On the other hand, in the Falicov-Kimball model, solving the impurity problem can be analytically carried out~\cite{Freericks1998,Freericks_rev2003}.
The $f$ electron number is given by $\means{n_i^f}=1+\frac{1}{2}e^{\beta(\varepsilon_f-\mu)}A_i^2$ with $A_i=\prod_n[1-U'{\cal G}_i(i\omega_n)]^{-1}$, and the self-energy is calculated from $\Sigma_i={\cal G}_i^{-1}-[G_i^{\rm imp}]^{-1}$, where $G_i^{\rm imp}$ is the impurity Green function given by $G_i^{\rm imp}=(1-\means{n_i^f}){\cal G}_i+\means{n_i^f}[{\cal G}_i^{-1}-U']^{-1}$.
Here, we assume the paramagnetic system and the Green functions and self-energy does not depend on spin.
We iterate the above procedure until the self-energy converges for all $\omega_n$ at each site.

The magnetic susceptibility for conduction ($f$) electrons defined by $\chi^{cc(ff)}=\frac{1}{N_s}\sum_{ij}\int_0^\beta d\tau\left[\means{M_i^{c(f)}(\tau)M_j^{c(f)}(0)}-\means{M_i^{c(f)}(\tau)}\means{M_j^{c(f)}(0)}\right]$ with $M_i^{c(f)}(\tau)=e^{\tau({\cal H}-\mu N)}\frac{1}{2}(n_{i\uparrow}^{c(f)}-n_{i\downarrow}^{c(f)}) e^{-\tau({\cal H}-\mu N)}$ is also evaluated from $\Sigma$ and $G^{\rm latt}$~\cite{Freericks1998,Freericks_rev2003}.
The $f$-electron susceptibility is simply given by $\chi^{ff}=\frac{1}{N_s}\sum_i\frac{\means{n_i^f}}{4T}$ and $\chi^{cf}=0$.
On the other hand, the magnetic susceptibility for the conduction electrons is more complicated due to the presence of the intersite correlations.
This is given by $\chi^{cc}=\frac{1}{2N_s}\sum_{ij}T\sum_n\chi_{ij}^{cc}(i\omega_n)$ with $\chi_{ij}^{cc}(i\omega_n)=[\chi^0(I+\Gamma \chi^0)^{-1}]_{ij}$,
where the bare susceptibility and vertex function are $\chi_0^{cc}=\frac{1}{2N_s}\sum_{ij}T\sum_n\chi_{ij}^{0}(i\omega_n)$ with $\chi_{ij}^0(i\omega_n)=-G_{ij}^{\rm latt} G_{ji}^{\rm latt}$ and $\Gamma_{ij}(i\omega_n)=\frac{\partial \Sigma_i}{\partial G_{ii}^{\rm latt}}\delta_{ij}=\frac{\Sigma_i(U'-\Sigma_i)}{1+G_{ii}^{\rm latt}(2\Sigma_i-U')}\delta_{ij}$, respectively.
In the following, we set the unit of the energy to the half-bandwidth $W$.

\section{Distribution of $f$-electron occupancy in Penrose lattice}\label{sec:distr-f-electr}

Before showing the numerical results for the Penrose lattice,
we briefly touch the DMFT results on the Bethe lattice,
whose density of states (DOS) is shown in Fig.~\ref{bethe}(a).
Figure~\ref{bethe}(b) shows the $f$ electron number $\means{n^f}$ at $U'/W=2$ and $n=1.9$,
where $W$ is the half bandwidth.
$\means{n^f}$ monotonically increases with decreasing $\varepsilon_f$ as expected.
At lower temperatures, there appears a flat region with $\means{n^f}\sim 0.1$
around $\varepsilon_f=-1$.
We refer to this as the intermediate valence region.
At $T/W=0.025$, we find a jump singularity in $\mean{n^f}$ around $\epsilon_f/W=-1.5$.
This suggests a first-order valence transition from the intermediate valence state
to the almost commensurate valence state with $\means{n^f}\sim 1$.

\begin{figure}[t]
 \begin{center}
  \includegraphics[width=0.8\columnwidth,clip]{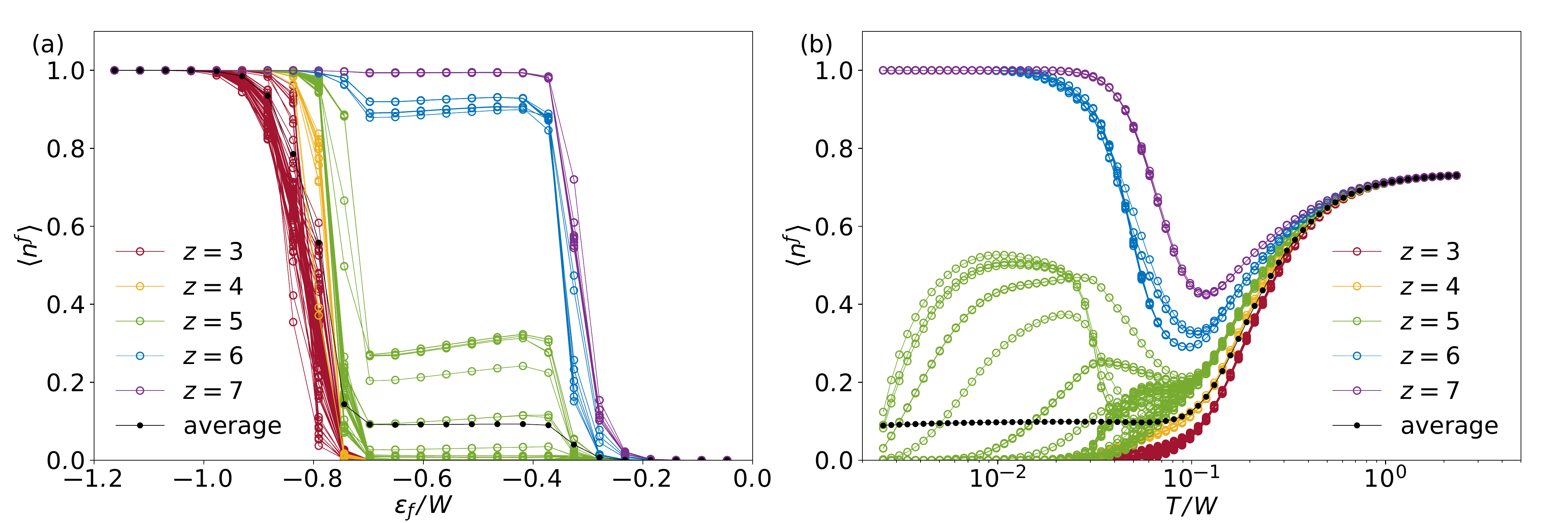}
  \caption{
(a) Site-dependent $f$ electron numbers as a function of $\varepsilon_f$ at $T/W=0.0233$ and (b) those as a function of temperature at $\varepsilon_f/W=-0.465$ in the Penrose lattice. Here, $U'$ is fixed to $U'/W=1.16$.
They are classified by the coordination number $z$.
}
  \label{nf}
 \end{center}
\end{figure}


Now, we consider the case of the Penrose lattice without translational symmetry,
whose half bandwidth is given by $W\simeq 4.3t$ as shown in Fig.~\ref{penrose}(b).
In the case, the $f$ electron number explicitly depends on the site position.
In Fig.~\ref{nf}(a), we present the $\varepsilon_f$ dependences of $\means{n_i^f}$
for all sites on the Penrose lattice.
The site average of $\means{n_i^f}$ is shown by the black filled circles.
In contrast to the case of the Bethe lattice, we do not find any anomalies corresponding to valence transitions at the lowest temperature we have performed.
More importantly, the $\varepsilon_f$ dependences of $\means{n_i^f}$ are roughly classified by the coordination number of their sites as shown Fig.~\ref{nf}(a).
This feature originates from the itinerant properties of the conduction electrons and has been discussed in other models with local interactions on the Penrose lattice, such as the Hubbard and extended Anderson lattice models~\cite{Takemori2015,Takemura2015,Shinzaki2016}.
In the intermediate valence regime around $\varepsilon_f/W = -0.6$, the site dependence of $\means{n_i^f}$ is observed remarkably.
$\means{n_i^f}$ at the sites with $z=6$ and $7$ take almost $1$,
while those with $z=3$ and $4$ vanish in this region.
We also find the large site dependence within the sites with $z=5$ shown in Fig.~\ref{nf}(a).

To clarify the more detailed nature presented in the intermediate valence regime, we show the temperature dependence of the $\means{n^f}$ at $\varepsilon_f/W=-0.465$ in Fig.~\ref{nf}(b).
At high temperatures, $\means{n_i^f}$ hardly depends on its site position,
and the quasiperiodic structure play a minor role in the electronic state.
With decreasing temperature ($T/W\sim 0.2$),
the variance of $\means{n_i^f}$ for the site position increases but this almost depends on the coordination number, namely the local structure including its neighboring sites.
Further decrease of temperature below $T/W\simeq 0.1$ brings about
more complicated site dependence.
In particular, we find that $f$-electron number for the lattice sites with $z=5$
are distributed in the wide range $0<\mean{n^f}\lesssim 0.5$.
Nevertheless, all $\means{n_i^f}$ for $z=5$ approaches zero
and the $f$ electron number depends only on the coordination number $z$ in the zero temperature limit.

\begin{figure}[t]
 \begin{center}
  \includegraphics[width=0.8\columnwidth,clip]{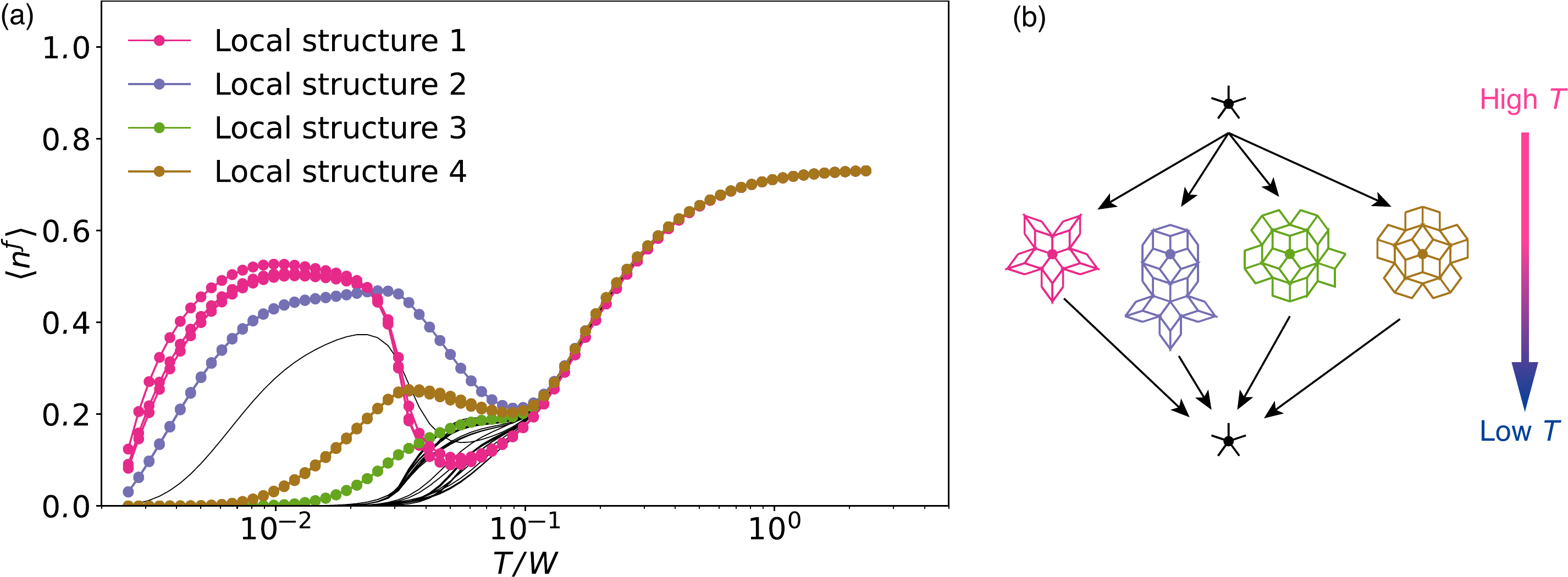}
  \caption{
(a) Site-dependent $f$ electron numbers for the site with $z=5$.
Each colored symbol represents $\means{n_i^f}$ on a particular site with the different local structure shown by the corresponding colors in (b).
}
  \label{figz5}
 \end{center}
\end{figure}

Figure~\ref{figz5}(a) shows the temperature dependence of the $f$ electron numbers for the sites with $z=5$,
which is extracted from Fig.~\ref{nf}(b).
In the Penrose lattice, similar local lattice structures appear with a certain density,
which is one of the most striking features of the quasiperiodic lattice.
Here, we focus on the four local structures with $z=5$ shown in Fig.~\ref{figz5}(b).
Figure~\ref{figz5}(a) shows the data on the site with these local structures using the corresponding colors.
This result indicates that the site dependence of $\means{n^f}$ around $T/W=0.01$ is classified by the wider range of lattice structure surrounding its site beyond the nearest neighbor one.
On the other hand, this feature for $z=5$ is expected to disappear at zero temperature.
The present results suggest that the wider range of lattice structure affects $\means{n^f}$ only at intermediate temperatures while the neighboring structure is relevant at zero and high temperatures.
This temperature dependence is schematically depicted in Fig.~\ref{figz5}(b).

\section{Temperature dependence of susceptibility}\label{sec:temp-depend-susc}
\begin{figure}[t]
 \begin{center}
  \includegraphics[width=0.8\columnwidth,clip]{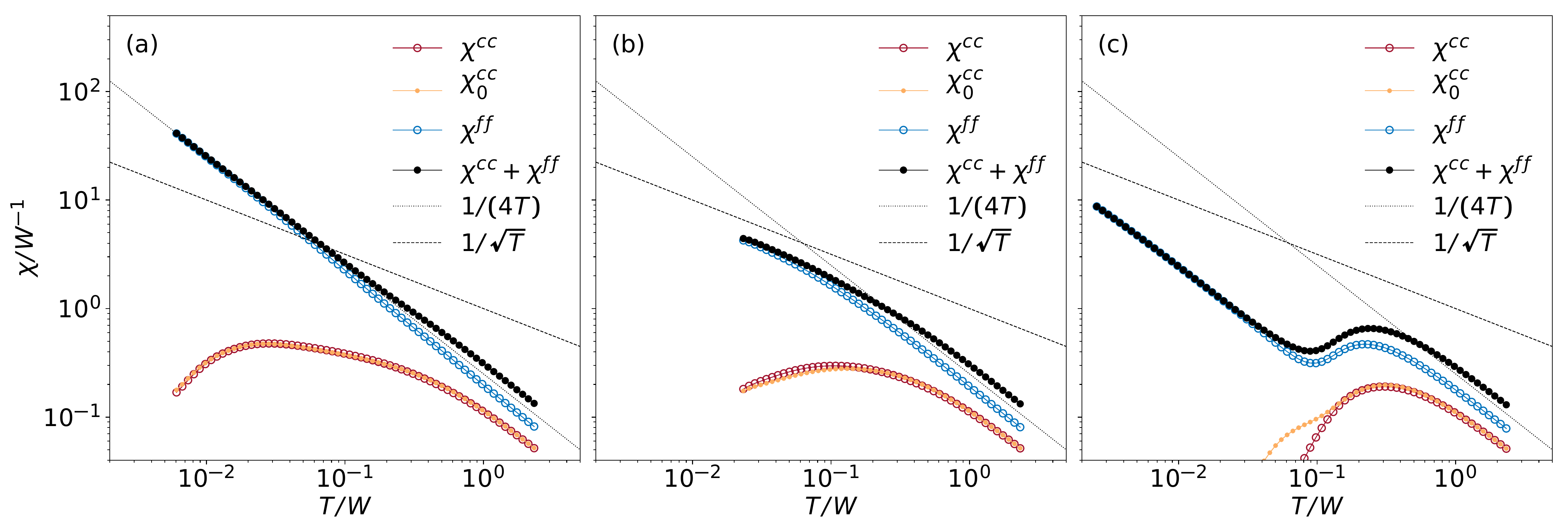}
  \caption{
Temperature dependence of the susceptibilities for the conduction and $f$ electrons at (a) $\varepsilon_f/W=-0.930$, (b) $\varepsilon_f/W=-0.767$, and (c) $\varepsilon_f/W=-0.465$.
The bare susceptibility $\chi_0^{cc}$ is also shown in each figure.
}
  \label{chi}
 \end{center}
\end{figure}

Finally, we discuss the magnetic response in the FKM on the Penrose lattice.
It was confirmed that the susceptibility is almost independent of the cluster size in a similar model~\cite{shinzaki2017cluster}.
Figure~\ref{chi} shows the temperature dependences of the susceptibilities $\chi^{cc}$ and $\chi^{ff}$ for the conduction and $f$ electrons, respectively.
Note that $\chi^{cc}$ includes intersite dynamical correlations within DMFT.
At the high temperature limit, $\chi^{cc}$ and $\chi^{ff}$ obey the Curie law.
With decreasing temperature, $\chi^{cc}$ increases but changes to decrease around $T/W=0.1$, which is commonly seen in the different $\varepsilon_f$ shown in Figs.~\ref{chi}(a)--\ref{chi}(c).
Note that the deviation of $\chi^{cc}$ from the bare susceptibility $\chi_0^{cc}$ is enhanced with increasing $\varepsilon_f$ at low temperatures.

On the other hand, $\chi^{ff}$ exhibits distinct temperature dependences for the different $\varepsilon_f$.
In the case with the deep $f$ electron level, $\chi^{ff}$ monotonically increases
with decreasing temperature, as shown in Fig.~\ref{chi}(a).
We then find asymptotic behavior $1/(4T)$ since $f$ electron level is filled at zero temperature.
In the intermediate valence regime ($\epsilon_f/W=-0.465$),
nonmonotonic behavior is exhibited, as shown in Fig.~\ref{chi}(c).
The decrease of $\chi_{ff}$ around $T/W\sim 0.1$ originates from
the monotonic decrease of the site average of $\means{n_i^f}$, as shown in Fig.~\ref{nf}(b).
Between the intermediate valence state and that with $\means{n^f}\sim 1$,
intriguing behavior appears.
Figure~\ref{chi}(b) shows that $\chi^{ff}$ obeys $1/\sqrt{T}$ at lower temperatures, which is also observed in the extended Anderson lattice model on the Penrose lattice~\cite{Shinzaki2016}.
Note that lower-temperature data could not be calculated because of the bad convergence in the DMFT scheme.
At $\varepsilon_f/W=-0.767$, the large site dependence of $\means{n_i^f}$ due to the Penrose lattice is seen in Fig.~\ref{nf}(a).
This might be an origin of the peculiar behavior of $\chi^{ff}$ in the competing region between the intermediate and commensurate valence regimes.

\section{Summary}\label{sec:summary}

We investigated the FKM on the two-dimensional Penrose lattice using RDMFT.
While the first-order valence transition exists in the FKM on the Bethe lattice, this is suppressed in the Penrose lattice.
Owing to the absence of the translational symmetry,
the $f$ electron number explicitly depends on the lattice site and is roughly classified by the coordination number, which may lead to the suppression of the valence transition.
In the intermediate temperature region, this is affected by the wider range of the lattice structure surrounding its site but the short range structure is only relevant at the lowest temperature.
We also calculated the magnetic susceptibility and found the peculiar temperature dependence, which might be due to the absence of the translational symmetry intrinsic in the quasiperiodic Penrose lattice.

\begin{acknowledgements}
This work is supported by Grant-in-Aid for Scientific Research from
JSPS, KAKENHI Grant Nos. JP16K17747(J.N.) and JP17K05536 (A.K.).
Parts of the numerical calculations were performed
in the supercomputing systems in ISSP, the University of Tokyo.
\end{acknowledgements}

\bibliographystyle{spphys}       
\bibliography{./refs}   

\end{document}